# Applying UML and MDA to Real Systems Design

Ian Oliver
Nokia Research Center, Finland
ian.oliver@nokia.com

## 1 UML and MDA

Traditionally system design has been made from a black box/functionality only perspective which forces the developer to concentrate on how the functionality can be decomposed and recomposed into so called components. While this technique is well established and well known it does suffer from some drawbacks; namely that the systems produced can often be forced into certain, incompatible architectures, difficult to maintain or reuse and the code itself difficult to debug. Now that ideas such as the OMG's Model Based Architecture (MDA) or Model Based Engineering (MBE)[1] and the ubiquitous modelling language UML are being used (allegedly) and desired we face a number of challenges to existing techniques.

When working with the UML, one **must** take into consideration **object orientation**. The UML is a language for expressing systems (or whatever) in terms of object oriented concepts and its meta-model and its semantics make this explicit. Object orientation, unlike functional based approaches makes *both* functionality *and* data first-class modelling elements. Whenever anything is specified in UML, that modelling element is either based on the notion of a class or is directly related to a class. Some methods appear to adhere to this but fail to use classes in this way by assuming the existence of a 'global' system and then just using classes as data elements. Effectively the UML equivalent of programming Fortran in C++.

Use case based development makes this situation worse by only concentrating only on behaviour and then also by not by integrating with the class-object concepts of OO. Use cases are often 'realised' as sequence diagrams, where the objects are never shown nor specified in a class diagram so that analysis of how the objects may be correctly connected or related is never achieved. In this approach objects themselves are just the objectification of functions.

The result of this is that systems developed using use cases suffer from poor or non existence object orientation, little or impossible to achieve reuse, every object in the system can communicate and be related to every other object in the system (coupling tends to be very high if not total) and that if a class structure is reverse engineered then it is often the case that most classes contain a single function which is often redefined in very deep inheritance hierarchies. Inheritance here often gets used as a development mechanism rather than the taxonomy mechanism it really is.

If use cases are to be used then they must either be as very high level requirements which are then subjected to *proper* OO analysis[2] and/or they are later used as tests to the system (cf: Model Based Testing). Use cases here are being used as the basis of scenarios that the customer believes they want; they can be thought of as scripts or constraints in the model checking sense. There is almost never a one-to-one mapping between the use cases and the functionality of the system that is finally constructed, just that the system is capable of providing the services or functionality required to enact the described scenario.

In an OO based system, the global behaviour or functionality is **emergent** from the particular collaborations and configurations of objects and their relationships rather than being specified explicitly for the whole system. Use cases in this respect are preferably used as (high level) tests to the model rather than first-class development artifacts.

The Model Driven Architecture is based around the concepts of models and transformations (or mappings) between those models. A transformation is a kind of model which contains information about the platform onto which you are mapping its source model(s). In a more generic scenario, a model may actually be a structure of models and a transformation a generic engine that takes a model of a platform as its parameter. The source models of a transformation are known as platform independent models (PIMs) and the target models as platform specific models (PSMs). These terms are *relative* to a given platform - given any model one can not state whether it is platform independent or platform specific without a second model related to it by one or more transformations.

What the MDA is trying to achieve is the axiomatisation of development and architectural principles; the knowledge held by engineers on solving certain problems is embodied inside the transformations. The situation however is no different from the ideas of compilation of $3^{rd}$ generation languages into machine code - transformation engines in some MDA tools are known as model compilers.

The MDA as defined by the OMG is based around that models are expressed in the UML or one of its profiles and that their meta-models are expressed in the Meta Object Facility or MOF. Transformations then act on the meta-model level transforming the concepts in one meta-model to concepts in another meta-model. However, as MOF can be considered a form of graphical BNF then any language's meta-model can be expressed in this form. Note that transformations operate at a **semantic level** rather than syntactic and may take a large amount of contextual or platform information into consideration. Compare this with the generally poor attempts at automatic code generation seen in many tools which seemingly equate the notion of a class at any abstraction level with that of a class in some 3GL programming language. A syntactic transformation implies that no change of abstraction level is made between the input and output of the transformation and effectively the input and output are the same but just expressed in different formalisms with the same semantics; this is commonly seen with say, SDL and C or UML or Java/C++.

---

[1] we use MDA in this article but the terms are generally interchangeable

[2] Uses cases are not object oriented!

## 2 Application in System Development

To correctly apply UML/MDA one must have a much greater understanding and adherence to the various levels of abstraction that are possible, a well defined separation of concerns and a process/method that actively supports modelling.

Abstraction relates to what information is required in models and how these models relate to each other. One of the problems with UML is that it is often used at the same level of abstraction as the implementation. There are very few tools that allow much further abstraction in a meaningful way. This is then coupled with areas where OO is unknown, for example hardware or protocol development, which gives way to concerns about the models not containing enough information or being too abstract for meaningful development. UML is designed as an extensible language which can be tailored for particular domains, it is in the system domain and the hardware aspects, that currently UML is particularly lacking.

MDA requires that, ideally, for each particular aspect or concern and at each level of abstraction a model is constructed. This separation of concerns, while good engineering practise, is rarely seen. Deciding which aspects are necessary for successful system development is difficult as is maintaining these concerns. At each abstraction level a well defined set of tests must be performed upon this system and maintained as the 'system models' are developed. At minimum one must have a separation between the domain of the system (what the system is) and the potential platforms (or architectures) onto which the system may be mapped. Separation of 'domain' and 'platform' is the key to success here and avoiding polluting either model with information from the other.

The language of these models varies depending upon abstraction level and aspect. There do exist numerous UML profiles for this work and in particular we note UML for QoS and Fault Tolerance, UML Profile for Schedulablity, Performance and Time, the UML Testing Profile, the Systems Modelling Language (SysML) and the UML for Communicating Systems (ETSI). Unfortunately these languages lack overall coherence, unless you count the core UML they are based upon, and lack of tool support. The other major problem here is that the concepts in these languages are often ambiguous even at a platform independent or generic level, for example, "what is a process?"

UML2 is not going to solve any problems here as to the casual user nothing has changed although there are new diagram types but less information on the semantics of these diagrams and once again how the diagrams and their elements relate to each other in the model. Classes are classes regardless of whether UML1.x or 2.x is used and their meaning context dependent.

This leads to then methodology (meaning the combination of methods and process). While we have languages and concepts for expressing artifacts and modelling elements relevant to systems development we are sorely lacking in consistent, well defined methods for the utilisation of these languages. As yet, while the MDA provides us with a methodological structure we are still missing how the modelling languages fit into that structure. A well defined set of semantic transformations do not exist for manipulating those models at their various levels of abstractions; those transformation that do exist are often syntactic in nature and do not effectively take into consideration the mapping via a platform into a more platform specific model.

Much of the methodology work required is deciding which languages (UML profile + extensions) to use at which level of abstraction and how each level of abstraction relates to each other. Certainly a core, basic language must be used at higher level of abstraction, eg: plain UML with each level progressively adding more information and changing the language. This change is dictated somewhat by the chosen platforms for mapping and thus the target implementation.

Finally if one is to model then one must have a reason about why those models are being produced. If a model can not be tested somehow then there is little point in producing that model - models must convey information to the users of those models. Testing here can mean metrics, validations (simulation, animation etc), verification (proof, model checking) and so on.

## 3 Conclusion

The problems faced by UML and MDA in systems development are broadly the same as in any other development area. The main problem we have encountered faced is that OO is not common in this area and that the UML is being misunderstood and not being applied correctly - the wrong types of analysis are being made (functional vs OO). Often it is just the case that the models do not contain any relevant information and are not being used to support the system under development - this is commonly seen with documentation oriented methods in which the documentation is more important than the actual product itself and thus the quality of that product. While UML has the potential to support other development paradigms it is still **inherently** an object oriented language *and* requires support from external processes and methods - this must be understood first before successful application.